\documentclass[sigconf]{acmart}

\AtBeginDocument{%
  \providecommand\BibTeX{{%
    \normalfont B\kern-0.5em{\scshape i\kern-0.25em b}\kern-0.8em\TeX}}}

\setcopyright{acmcopyright}
\copyrightyear{2023}
\acmYear{2023}

\acmConference[CHI '23]{Make sure to enter the correct
  conference title from your rights confirmation emai}{April 23--28,
  2023}{Hamburg, Germany}
\acmBooktitle{CHI '23: ACM CHI Conference on Human Factors in Computing Systems,
 April 23--28, 2023, Hamburg, Germany} 

\begin{document}

\title{Generative AI for Product Design: Getting the Right Design and the Design Right}

\author{Matthew K. Hong, Shabnam Hakimi, Yan-Ying Chen, Heishiro Toyoda, Charlene Wu, Matt Klenk}
\email{{matt.hong, shabnam.hakimi, yan-ying.chen, heishiro.toyoda, charlene.wu, matt.klenk}@tri.global}
\affiliation{%
  \institution{Toyota Research Institute}
  \streetaddress{4440 El Camino Real}
  \city{Los Altos}
  \state{CA}
  \country{USA}
  \postcode{94022}
}

\renewcommand{\shortauthors}{Matthew and Matthew Klenk, et al.}

\begin{abstract}
Generative AI (GenAI) models excel in their ability to recognize patterns in existing data and generate new and unexpected content. Recent advances have motivated applications of GenAI tools (e.g., Stable Diffusion, ChatGPT) to professional practice across industries, including product design. While these generative capabilities may seem enticing on the surface, certain barriers limit their practical application for real-world use in industry settings. In this position paper, we articulate and situate these barriers within two phases of the product design process, namely \textit{getting the right design} and \textit{getting the design right}, and propose a research agenda to stimulate discussions around opportunities for realizing the full potential of GenAI tools in product design.
\end{abstract}



\maketitle

\section{Introduction}
The publication of the seminal paper, ``Attention is all you need'', by Vaswani et al. \cite{vaswani2017attention} in 2017 marked an important milestone\footnote{Recent advances in generative AI are often attributed to Vaswani et al.'s paper, which describes a simple self-attention based Transformer architecture that significantly increases the quality of language models while reducing model training time.} in AI research, fueling an arms race toward the development of large scale language models (LLM's). The emergence of LLM's represent an inflection point in model performance where they begin to outperform humans in many benchmark tasks, including speech and image recognition, language understanding, among others\cite{sevilla2022compute,wei2022emergent}. 

These advances in LLM's provide the backbone for Generative AI (GenAI) models and recent explosion of creativity support tools\cite{Insights2023-fi} that present promising opportunities to augment human creativity in real-world professional practice. Brands such as Nike and Heinz and creative ad agencies already use text-to-image GenAI tools to create product placement advertisements\cite{Bonilla2022-si}. Furthermore, an artwork created with Midjourney won Colorado State Fair's fine art competition, stirring debate over its legitimacy\cite{Roose2022-rh}. While exciting, it is unknown at this point if these tools are improving human performance, if so, in what ways, and what interfaces and interactions will best support professionals. 

Designing digital and physical products is a labor-intensive endeavor. At its core is an iterative process of idea elaboration and reduction, where each step adds significant cognitive, psychological, and physical burden for the designer. Innovative design practitioners emphasize the importance of \textit{getting the right design} prior to \textit{getting the design right} and then investing in implementing fully functional systems\cite{buxton2010sketching}. The separation of these two approaches allows the designer to prioritize finding the \textit{global maxima} first instead of optimizing too early and settling for suboptimal solutions. In theory, GenAI may address this problem by reducing the burden of elaboration and by enticing the designer in new directions at a low cost. In either case, working with GenAI models is significantly different from industrial design practice and therefore, significant support will be required.

In this position paper, we contribute a discussion around HCI challenges to applying GenAI in each of these phases. In \textit{getting the right design}, we identify a potential tradeoff between inspiration and the efficiency at generating high-fidelity designs along with challenges in the current interaction modalities of these tools and designer expectations. In \textit{getting the design right}, we identify a human-centered challenge of achieving common ground with the GenAI model and a technical challenge of how different models of consumer preference should influence the process.

\section{Getting the Right Design}
Design is a wicked problem which has no single definitive solution---there is only constant tension between elaboration of a possible solution and generating new solutions in another part of the design space. Combating such uncertainty in design requires a thorough investigation of the design space by enumerating meaningfully distinct choices of ideas, and adding elaborations to these ideas to identify the \textit{global maxima}. By design, the generative capabilities of GenAI should support designers in accelerating their exploration of the design space, yet there are important considerations that challenge their use for professional practice.


\subsection{High-fidelity design or inspiration?}
The ease of which GenAI tools generate photorealistic images, while efficient, also creates new problems for design. Moving to high-fidelity so quickly could actually have a negative impact on designer creativity. For example, showing a high-fidelity solution so early in the process may induce \emph{design fixation} \cite{youmans2014design}, the process in which designers continue to offer variations on an existing design without considering alternative solutions. The HCI and design disciplines have emphasized the gradual evolution of designs from low to high fidelity with evaluation interspersed throughout iterative steps in the design cycle. The additional time spent moving from low-fidelity to high-fidelity may result in important learnings that may lead to creative breakthroughs.

This makes the compelling case to treat GenAI as tools for inspiration, instead of outright design, in this early stage of the design process. For instance, many product designers draw inspiration from community-driven visual content curation websites such as Pinterest and Behance, in order to generate mood boards that help steer their creative process. However, because these websites are optimized to attract user engagement, the displayed content may prematurely constrain the boundaries from which designers draw their inspiration. GenAI can fill this gap by providing computational means of generating alternative sources of inspiration\cite{sbai2018design}

\subsection{Increasing idea diversity}
While many generative AI tools and services already focus on providing augmentations and variations of human-created visual content, they often create variations on visual style than the ideas presented in the image. Navigating the design space requires out-of-the-box thinking, which comes from the process of elaborating on `meaningfully distinct' ideas. These ideas, however, are susceptible to, and often influenced by the designers' own intuition, experience and biases about the topic. 

To this end, future development of GenAI tools should consider computational means for increasing idea diversity by offering ideas that are visually and semantically distinct from each other, and create appropriate mechanisms for users to control the desired level of diversity to prevent significant deviations from each idea. Idea diversity can also come from systematic explorations of existing designs across product categories\cite{hope2022scaling} and domains\cite{kang2022augmenting}. To make GenAI tools useful for product design, we should explore opportunities to integrate our knowledge of conceptual mappings in natural language--that are sensitive to different social and cultural contexts--as well as human-made products that embody some knowledge of the problem and design space.

\subsection{Prompt engineering challenge}
Text-to-image translation systems offer designers the ability to translate envisioned concepts into photo-realistic design artifacts in just a matter of seconds. However, adopting prompt engineering-based design in business practice is challenging because of the cognitively difficult task of translating designers' visual concepts into text expressions. This translation involves clearly articulating intended meanings and remembering specific design ontology (e.g., surrealism, extreme close-up shot) that is recognized by the generative model. This problem leads to text prompts that result in image depictions that are inconsistent with the designer's intended visual concept, or vice versa, thus adding significant time to iteratively refine the prompt until the desired outcome is achieved\cite{hutchinson2022underspecification}. In one example, using DALL-E 2's GenAI system, it took Nestlé more than 1,000 text prompts, followed by human evaluation of each, to arrive at its final product placement advertisement for its yogurt product by re-rendering Johannes Vermeer's painting ``The Milkmaid''\cite{Bonilla2022-si}.

By interactively prompting users to specify and correct these details (either through language or imagery), GenAI systems can aid the user in iteratively improving the scene in fewer input-output loops. Only then can designers begin to consider adding variations while retaining control over specific design elements.

\section{Getting the Design Right}
While the responsibility of \textit{getting the right design} falls mostly on the designer, \textit{getting the design right} requires a concerted effort among design, engineering, marketing, and other invested stakeholders. Once a design problem is defined, the product team must make continuous iterative refinements to an idea to narrow down the design space. Doing so requires making decisions on concepts to align with the goals set forth by the product team. These goals are informed by the team's analysis of engineering requirements, rigorous usability testing, and measuring consumer reactions to the product, etc. Here, we focus on capturing consumer preferences, a challenging prospect that adds significant cost burden and uncertainty in \textit{getting the design right}.

\subsection{Maintaining design goals}
As a company brings a new product to market, it will learn that there may be trade-offs between different functional requirements. For example, consumers may want a sturdy artifact that is also light and inexpensive. Throughout the product development cycle, engineers will make decisions that navigate these trade-offs typically without strong links to the information used in the requirements engineering process. Providing this information to decision-makers in their workflow has the potential to significantly improve the resulting design while reducing lead time to bring new products to market. An important challenge here is how to represent the consumer's desires begin as under-specified requirements for later refinement and communication.

Consumers do not care solely about the functional capabilities of the product. Aesthetic considerations drive many purchase decisions. As the design moves from concept to production, many engineering decisions will impact the aesthetic of the resulting artifact. By maintaining the designer's views of the consumer's preferences, these engineering changes can be evaluated for their aesthetic impacts without consulting expensive design panels. The larger challenge exists in representing the tension between functional and aesthetic requirements and teaching GenAI models to generate results that consider the combination of these constraints.

\subsection{Multi-modal representations of consumer preference}
Predicting how the public will respond to new products is a central problem for all firms. Vast consumer preference data are expressed and captured in the form of unstructured text as well as behavioral, and physiological signals. Data are analyzed in service of creating bespoke personas that represent consumer preferences, including the target customers' goals, demographic and socioeconomic background, lifestyle, values, etc. These personas play an important role in the design process, particularly to help ground the design in the context of data. Text-to-text GenAI tools such as ChatGPT or other LLM-based variants can provide early capabilities to generate personas that embody recent trends in consumer preferences and even play the role of the persona by responding to users' natural language queries. 

While there is merit in harnessing textual representations of consumer preference, there is inherent risk in treating LLM-based personas as proxy for a target demographic. We don't fully understand whose opinions are reflected in the generated responses, and there is potential for representational harms---there is either too much or a lack of representation from demographic subgroups\cite{santurkar2023whose}.

We believe there is potential in considering other types of data, including behavioral and physiological signals. There is huge risk in not knowing how consumers will behave in the presence of the designed product and if there are certain underlying physiological signals that help predict their preference toward functional or aesthetic features of a product, especially since preferences over such features may be outside of a consumer's conscious awareness. 

For GenAI to be useful in grounding the process of \textit{getting the design right}, we call for a greater need for the broader HCI community to invest in efforts to generate behavioral\cite{barthet2022generative} as well as physiologically-based personas. For instance, the emerging field of neurodesign applies research findings from psychology and applied neurosciences to understand how people respond to various kinds of design and visual stimuli\cite{auernhammer2021neurodesign}. Cross-disciplinary efforts in this direction may yield fruitful outcomes for grounding product design.

\subsection{Shifting preferences with time}
The most frustrating aspect of industrial product design is that consumer preferences shift across demographics and time. Once we have consumer preference models underlying the functional and aesthetic requirements, it is necessary to understand that products are not designed for the average consumer. Instead, designers use personas and market segments to target their products. During design, the target market segments may be refined, additional segments may be considered, or new consumer preferences may be identified. By linking consumer preferences to functional and aesthetic considerations, we should be able to update them dynamically based on new information.

This requirement, however, challenges our use of frozen LLM's, which limits the designer's ability to learn new trends in the market. Retraining LLM's incurs significant costs, financially and environmentally, and may present practical barriers to consider their use for understanding quick changes in consumer preferences. One practical use of text-to-text GenAI tools is to understanding slow-changing trends while using complementary methods to understand emerging trends at the time when newly designed products will be introduced to the market.
\section{Conclusion}
We conclude this paper by bringing awareness to the legal and ethical risks that accompany the use of GenAI tools as they are trained on billions of intellectual property without giving proper credit to the original creators. Significant efforts are also needed in addressing the lack of diversity and inclusion in datasets used to train GenAI models. While we do not discuss these problems in this paper, we do believe that responsible use of GenAI will ease potential strain in our society as well as empower product design teams.

\bibliographystyle{ACM-Reference-Format}
\bibliography{main}

\end{document}